# Identification of Risk Extreme Values in a Time Series and Analysis with an Autoregressive Method

Application for Climate Risk Events

Gianluca Rosso

(GradStat, RSS the Royal Statistical Society, London – Corresponding Scholar, SIS Società Italiana di Statistica, Roma)
*Correspondence to: gianluca.rosso@sis-statistica.org*

ABSTRACT. *In this article there is no intention to repeat basic concepts about risk management, but we will try to define why often is usefull the time series analysis during the assessment of risks, and how is possible to compute a significative analysis using regression and autoregression. After some basic concepts about trend analysis, will be introduced some methods to identify peaks. This is often usefull when there is no need to use the full time series, because sometimes is more practical to focus only on the extremes. With a correct time series without "not-anomalous" data, the extremes time series are treated with a simply autoregression model. This drives to know if the time series has a correlation between periods, and how many periods could be considered lagged among them. We think that climate events frequently are lagged because the climate show a clear increasing tendency, and that climate risks are potentially increasing during the time. There will be no specific conclusion related with risk management, because the proposed solution with autoregression can be adapted to any time series analysis.*

1. INTRODUCTION.

In this article there is no intention to repeat basic concepts about risk management, but few words about the principles of ORM will be usefull to introduce the question: why we need to use (or if you prefer: is better to use) time series analysis? And why this is usefull for the most of operational risk, but particulary for risks related with climate events?
Risk assessment is substantially a quantitative analysis, that needs both mathematical and statistical approaches. Are used the scrupulousness of math, but also the probabilistic concept of stat. So there is a balance between what "is" and what "could be". Another direction from this balance is a psychology approach, well known as "psychology of loss". The mile stone of this approach is the loss aversion.

Loss aversion is a human characteristic that describes how people are intrinsically afraid of losses. Loss aversion unavoidably leads to risk aversion and a number of predictable behaviours in certain situations. Loss aversion and risk are intrinsically linked. Research into the psychological value drives to the weight that people give to probabilities. And here is again the statistical element described more over. Where the odds of an event are very small, people become almost completely indifferent to variations in levels of risk. In particular people are more heavily influenced (in terms of weighting of probabilities) if an event is described by using frequencies than by using standard indicators of probability of risk. Loss aversion and related biases are an important drivers of human decision making in many situations. On the one hand it can encourage extreme risk averse behaviour, and yet in a different situation it can lead to excessive risk taking.

These are heavy evaluation elements in order to assess extreme events, as in climate.

Traditionally time series are the basis of any climate analysis. There is only mathematical and statistical reasons? Yes, if you refers to letterature. But probably is possible to add an element in order to make stronger this approch. The element is the loss aversion and the tendency of people to have a bias in favour of the "status quo" because they are more concerned about losses than about future gains. A tendency that is related to "future discounting", a basic concept related with the "Giddens's paradox".

It states that in the course of day-to-day life a great number of risks, and climate risk are in that order, many will sit on their hands and do nothing of a concrete nature about them. Yet waiting until they become visible and acute before being stirred to serious action will, by definition, be too late.

The key of this situation is that if people have troubles to see forward, researchers must operate in order to let the people see backward. And from the backward analysis, the time series analysis, we can see then forward. This because analysis of time series can drives automatically to the analysis of trend. This is a natural approach to the study, and there is no other way forward.

When we look to natural extreme events we approach to an argument drammatically important for humanity. Social, economic and moral implications are so big that its study will be one of the most important challenge for the next decades.

For example, global annual temperatures in last ten years seems to be in average quite stable. But if we wide our view and we take a look to the last 30 years, we could have a point of view very different. Because we can see clearly the tendance: temperature is growing up. It's true that we can observe at least two periods in which temperature are stable, but data mustn't be treated as isolated elements. They must be analized as a time series.

It's the natural variability that could show every single period as a record, in positive or in negative. Every portion of our time series could be anomalous, and therefore could contain an extraordinary event.

As wrote in many papers, as said in many news, the problem is not only the amount of the event but also the frequency of occurrence.

There are many ways to analyze these data; in this article we propose a path to have a solution.

After some basic concepts about trend analysis, will be introduced some methods to identify peaks. This is often usefull when there is no need to use the full time series, because sometimes is more practical to focus only on the extremes. With a correct time series without "not-anomalous" data, the extremes time series are treated with a simply autoregression model. This brings to know if the time series has a correletion between periods, and how many periods could be considered lagged. We think that single climate events are lagged because the climate show a clear tendency and that climate risks are potentially increasing during the time.

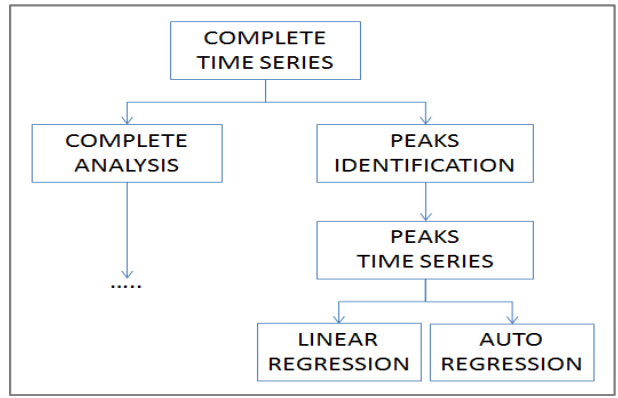

*Fig. 1.1*

## 2. TREND ANALYSIS.

At the basis of any analysis of extreme indicators lies the estimation of trends.
There are two definitions for "trend". In much of the statistical literature a trend is conceived of as that part of a series which changes relatively slowly (smoothly) over time. Viewed in terms of *prediction*, a trend is that part of the series which, when extrapolated, gives the clearest indication of the future long-term movements in the series.
In many situations these definitions will overlap. But not in all situations. In case of data following a *random walk*, the latter trend definition does not lead to a smooth curve. Typical examples of the first definition are splines, LOWESS smoothers and Binomial filters. Typical example of the second definition is the IRWtrend model in combination with the Kalman filter.
If we scan the climate literature on trend methods, an enormous amount of models arises. We found the following trend models or groups of models (without being complete): low pass filters (various binomial weights; with or without end point estimates), ARIMA models and variations (SARIMA, GARMA, ARFIMA), linear trend with OLS (ordinary least square), kernel smoothers, splines, the resistant (RES) method, Restricted Maximum Likelihood AR, based linear trends, trends in rare events by logistic regression, Bayesian trend models, simple Moving Averages, neural networks, Structural Time-series Models (STMs), smooth transition models, Multiple Regression models with higher order polynomials, exponential smoothing, Mann-Kendall tests for monotonic trends (with or without correction for serial correlations), trend tests against long-memory time series, robust regression trend lines (MM or LTS regression), Seidel-Lanzante trends incorporating abrupt changes, wavelets, Singular Spectrum Analysis (SSA), LOESS and LOWESS smoothing, Shiyatov corridor methods, Holmes double-detrending methods, piecewise linear fitting, Students t-test on sub-periods in time, extreme value theory with a time-varying location parameter and, last not but least, some form of expert judgment (drawing a trend "by hand"). However, the number of trend models applied to extreme indicators, appears to be much more limited.
The trend model almost exclusively applied, is the OLS straight line. This model has the advantage of being simple and generating uncertainty information for any trend difference [$\mu_t - \mu_s$] (indices "t" and "s" are arbitrary time points within the sample period).
The OLS regression model reads as

$$y_t = \mu_t + \varepsilon_t = a + b*t + \varepsilon_t \qquad (2.1)$$

with "a" the intercept, "b" the slope of the regression line and "ₜ a noise process.
Other examples of OLS linear trend fits can be found in linear trend estimation in combination with ARIMA models for the residuals. In the field of *disaster studies* OLS trends are the dominant method, albeit that the original data are log-transformed in most cases. Hu et al. (2012) apply Mann-Kendall tests with correction for serial correlation (no actual trend estimated in this approach).
Finally, some authors acknowledge that the use of a specific trend model, along with uncertainty analysis, may lead to deviating inferences on (significant) trend changes. Therefore, they chose to evaluate trends using *more than one trend model*. For example, Moberg and Jones apply two different trend models to the same data: the OLS trend model and the resistant (RES) model. Subsequently, they evaluate all their results with respect to these two trend models. Even more methods are evaluated by Young et al. (2011). They estimate five different trend models to 23-yr wind speed and wave height data and evaluate uncertainty information for each model (their supporting material).
We note that the application of more than one trend model to the same data has been published more often (not specifically for the evaluation of extremes).

3. IDENTIFING PEAKS.

There are two methods in identifying movements of the extreme value are taking the maximum value in a given period (called Block Maxima method) and retrieve the values that pass through a given threshold (called Peaks Over Threshold method).
A dataset of events is given, and it is possible to draw it with a scatterplot (Fig. 3.1).

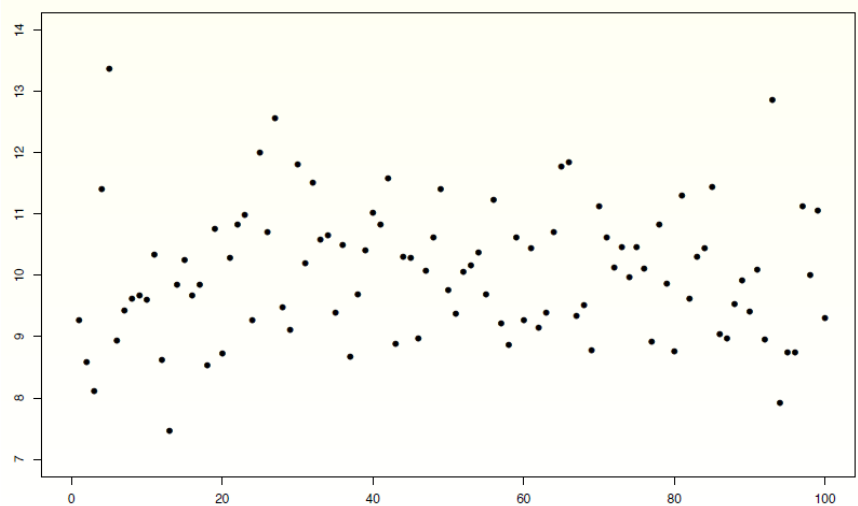

*Fig. 3.1. Source: E. Masiello, Univ. Lyon 1*

Generalized Extreme Value (GEV) distributionis a family of continuous distribution is built into the EVT to combine the Gumbel, Frechet, and Weibulldistribution, known as the extreme value distribution of type I, II, and III. On the GEV distribution, X is the random variable which has Probability Density Function (PDF) is as follows

$$f(x;\mu,\sigma,\xi) = \begin{cases} \dfrac{1}{\sigma}\left\{1+\xi\left(\dfrac{x-\mu}{\sigma}\right)\right\}^{-\frac{1}{\xi}-1} \exp\left\{-\left(1+\xi\left(\dfrac{x-\mu}{\sigma}\right)\right)^{-\frac{1}{\xi}}\right\}, & \xi \neq 0 \\ \\ \dfrac{1}{\sigma}\exp\left\{-\dfrac{x-\mu}{\sigma}\right\}\exp\left\{-\exp\left(-\dfrac{x-\mu}{\sigma}\right)\right\}, & \xi = 0 \end{cases}$$

(3.1)

where:

μ is a parameter of location;

σ is a parameter of scale;

ξ is a parameter of shape.

Shape parameter ξ determines the behavior of the tail distribution. Distribution type defined with $\xi = 0$, $\xi > 0$, and $\xi < 0$ and can be likened to the Gumbel, Frechet, and Weibull distribution.

One of the methods in identifying movements of the extreme value is Block Maxima, which identifies the extreme value based on the value of maximum observation data are grouped according to certain periods. In this method, observation data is divided into blocks in a certain period, example monthly, quarterly, semester, or year. Then for each block is determined the magnitude of observation data and the value is the maximum value of the extremes for each block and were used as the sample (Fig. 3.2).

The POT (Peak Over Threshold) method involves choosing some threshold value, collecting the extreme values above that threshold into a sample, and then drawing conclusions based on that sample. The extreme values are taken as the peak values between two distinct upcrossings. To account for variation in the particular time series chosen, it is beneficial to generate or record multiple data for the same length of time and then form the extreme value sample from the peaks over threshold of all the time series (fig. 3.3).

In our example is shown in Fig. 3.4.

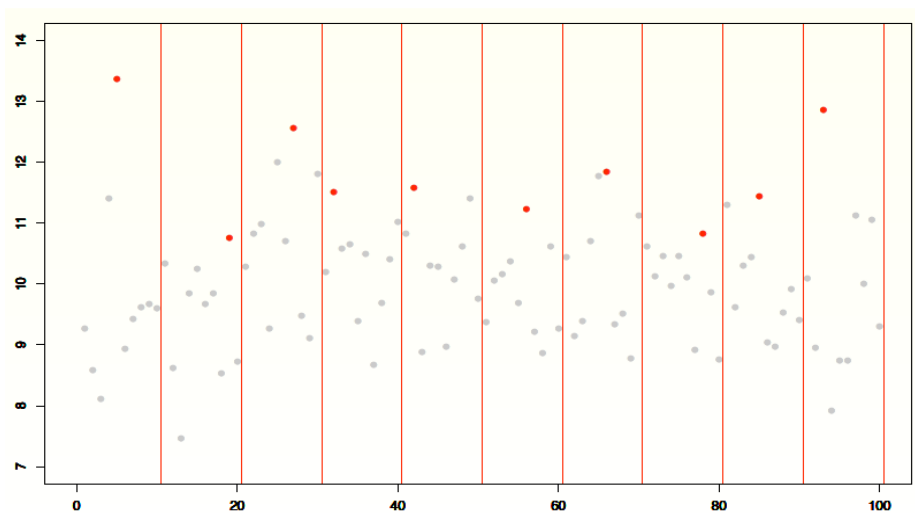

*Fig. 3.2. Source: E. Masiello, Univ. Lyon 1*

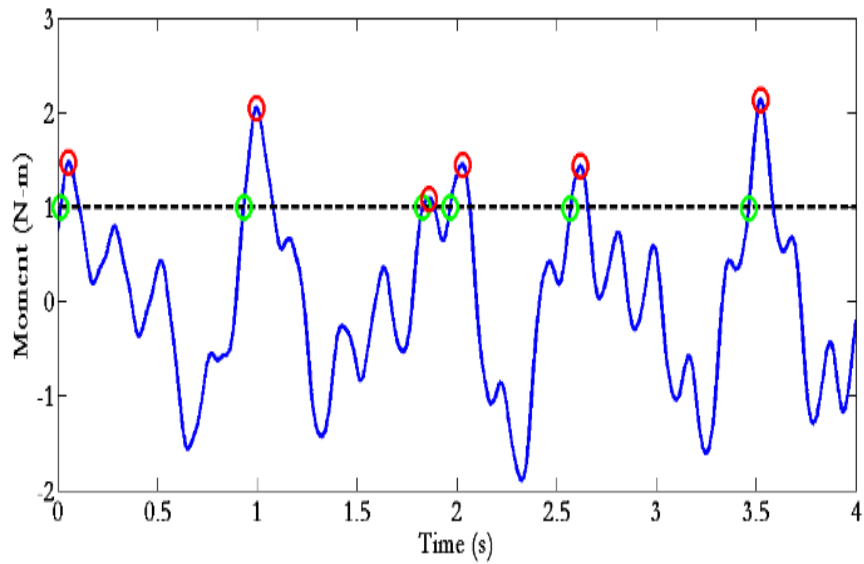

*Fig. 3.3. Example of time series with the upcrossings (green) and extreme values (red) marked for a threshold value. Source: J. Rinker, Peak-over-Threshold Method for Extreme Values.*

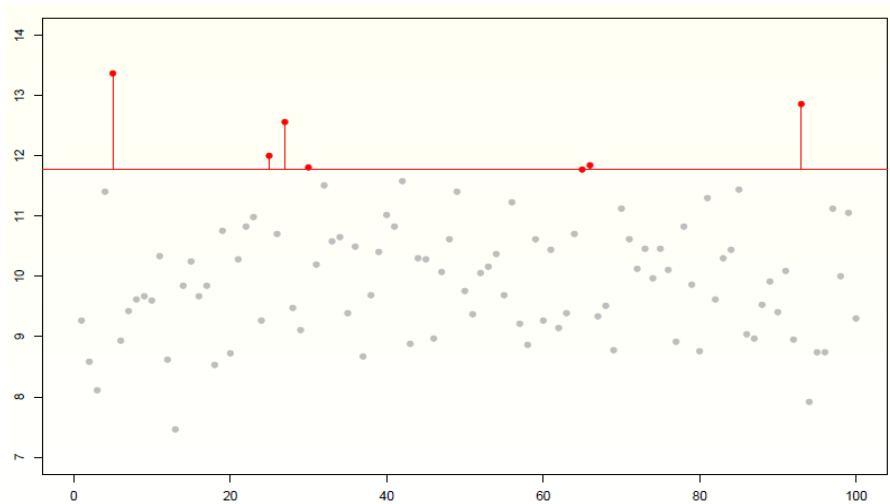

*Fig. 3.4. Source: E. Masiello, Univ. Lyon 1*

Let introduce an important relation between this method of analysis and the regression. This relation will be important for later consideration.
In the scatterplot above we have a distribution of items that has no particulary tendencys. Is a partial indefinite cloud of dots. But if we have a more definite trend of data, we should consider two approach.
With a constant threshold the graph is not much different from the other above (Fig. 3.5), but in a quantile regression the aspect of our analysis assume a different perspective (Fig. 3.6).

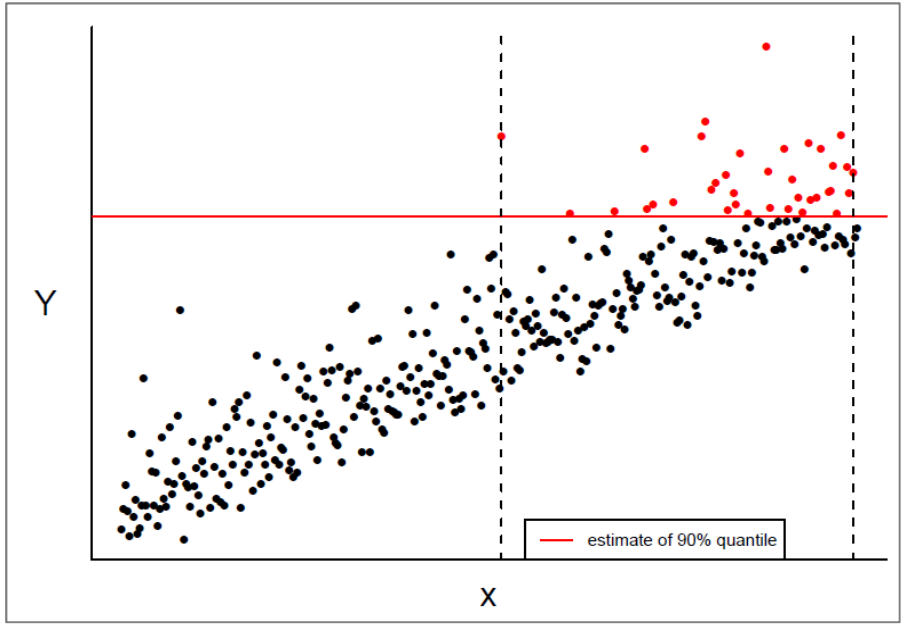

*Fig. 3.5.*

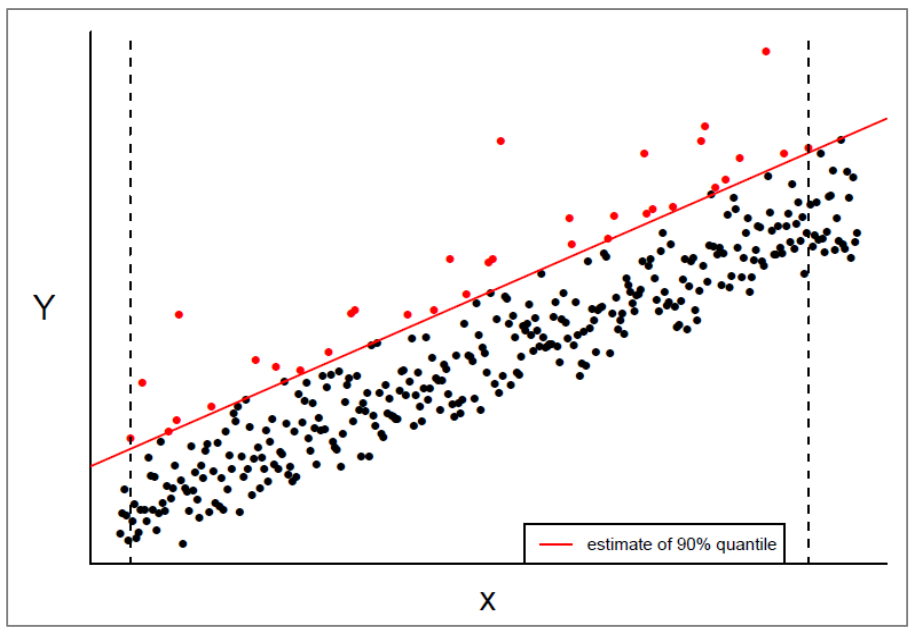

*Fig. 3.6.*

About the exceeding of the value over a threshold, is possible to calculate the probability of the loss during the time. We refer to a Keith Porter work[1]. This function is called risk curve.

---

[1] K. Porter, A Beginner's Guide to Fragility, Vulnerability, and Risk, University of Colorado, Denver CO USA.

$$R(\chi) = \int_{s=0}^{\infty} -(1 - P[X \leq \chi \mid S = s]) \frac{dG(s)}{ds} ds \tag{3.2}$$

where
- $X$ = uncertain degree of loss to an asset, such as the uncertain damage factor
- $x$ = a particular value of $X$
- $s$ = a particular value of the environmental excitation
- $R(x)$ = annual frequency with which loss of degree $x$ is exceeded
- $G(s)$ = the mean annual frequency of excitation exceeding intensity $s$
- $P[X \leq x \mid S = s]$ = cumulative distribution function of $X$ evaluated at $x$, given $s$. If $X$ is lognormally distributed at $S = s$

then

$$P[X \leq x | S = s] = \Phi\left(\frac{\ln(\frac{x}{\theta(s)})}{\beta(s)}\right) \tag{3.3}$$

where
- $\theta(s)$ = median vulnerability function, i.e., the value of the damage factor with 50% exceedance probability when the asset is exposed to excitation $s$
- $v(s)$ = coefficient of variation of vulnerability, i.e., the coefficient of variation of the damage factor of the asset exposed to excitation $s$
- $\beta(s)$ = logarithmic standard deviation of the vulnerability function, i.e., the standard deviation of the natural logarithm of the damage factor when the asset is exposed to excitation $s$.

About the environmental excitation, a correct definition we can give is that it it is usually the force, deformation, or other degree of loading to which the asset is subjected. In climate events, every measurable force is potentially an environmental excitation. About floods,
flood depth or momentum flux or a vector combination of the two at the perimeter of an asset such as a building qualify as excitation by the environment on the asset. The US Army Corps of Engineers uses these measures as inputs to their flood vulnerability functions.
If one has the mean vulnerability function $y(s)$ and coefficient of variation of loss as a function of excitation $v(s)$, to evaluate $\theta(s)$ and $\beta(s)$ using the following most general formulas

$$\beta = \sqrt{\ln(1 + v^2)} \tag{3.4}$$

$$\theta = \frac{\mu}{\sqrt{1+v^2}} \tag{3.5}$$

Equation *(3.2)* can be numerically integrated by

$$R(x) = \left(p_{i-1}(x) G_{i-1}\left(1 - exp(m_i \Delta s_i)\right) - \frac{\Delta p_i(x)}{\Delta s_i} G_{i-1}\left(exp(m_i \Delta s_i)\left(\Delta s_i - \frac{1}{m_i}\right) + \frac{1}{m_i}\right)\right) =$$
$$\sum_{i=1}^{n}(p_{i-1}(x) \cdot a_i - \Delta p_i(x) \cdot b_i) \tag{3.6}$$

from the *(3.3)*

$$p_i(x) = P[X \leq x | S = s_i] = \Phi\left(\frac{\ln(\frac{x}{\theta(s_i)})}{\beta(s_i)}\right) \qquad (3.7)$$

then

$$\Delta p_i(x) = p_i(x) - p_{i-1}(x) \qquad (3.8)$$

and

$$\Delta s_i = s_i - s_{i-1} \qquad (3.9)$$

$$m_i = \ln\left(\frac{G_i}{G_{i-1}}\right) / \Delta s_i \qquad (3.10)$$

for *i=1,2,…n*

4. REGRESSION.

Just a few words about regression, to introduce the autoregression model.
Linear regression rappresents an expected value estimation method, conditioned by a dependent variable Y, given values of others independent variables

$$X_1, \ldots, X_k : \mathbb{E}[Y | X_1, \ldots, X_k] \qquad (4.1)$$

These values are defined as regressors.
The linear regression model is

$$Y_i = \beta_o + \beta_1 X_i + u_i \qquad (4.2)$$

where:

- *i* change between observations, *i* = 1, ...., n;
- $Y_i$ is the dependent variable;
- $X_i$ is the independent variable or regressor;
- $\beta_0 + \beta_1 X$ is the regression line or population regression function;
- $\beta_0$ is the intercept value of the regression line;
- $\beta_1$ is the slope of the regression line;
- $u_i$ is the random error.

Extimated regression model is

$$\hat{Y}_i = \hat{\beta}_o + \hat{\beta}_1 X_i \qquad (4.3)$$

With

$$\hat{\beta}_1 = \frac{\sum(X_i - \overline{X})(Y_i - \overline{Y})}{\sum(X_i - \overline{X})^2} \qquad (4.4)$$

$$\hat{\beta}_0 = \overline{Y} - \hat{\beta}_1 \overline{X} \qquad (4.5)$$

In some cases the regression can be considered already sufficient enought to determine the trend. But sometimes the Mann-Kendall[2] test must be properly executed after the regression, that is the basis for the analysis. Shortly, we could say that linear regression is usefull to estimate tendency of the parameters, and Mann-Kendall test to estimate significance of the tendence. The choice about only regression or regression plus Mann-Kendall test depends mainly on the modeling of the investigation to be carried out, and then some conditions such as the availability of data, timing, costs.

---

[2] The Mann-Kendall (M-K) Test is a simple test for trend. Mann-Kendall is a non-parametric test and as such, it is not dependent upon: The magnitude of data, Assumptions of distribution (does not have to have a normal / bell shape distribution), Missing data or Irregularly spaced monitoring periods. Mann-Kendall assesses whether a time-ordered data set exhibits an increasing or decreasing trend, within a predetermined level of significance.

Using one of the test, or using a combination of both, five types of trend tests are possible (as presented in the following table)

|  | Not Adjusted for X | Adjusted for X |
|---|---|---|
| Nonparametric | Mann-Kendall trend test on Y | Mann-Kendall trend test on residuals R from LOWESS of Y on X |
| Mixed | ---- | Mann-Kendall trend test on residuals R from regression of Y on X |
| Parametric | Regression of Y on T | Regression of Y on X and T |

The table uses the following notation:
- Y = the random response variable of interest in the trend test,
- X = an exogenous variable expected to affect the value of Y,
- R = the residuals from a regression or LOWESS of Y versus X, and
- T = time (often expressed in years).

Mann-Kendall is frequently used for analysis on climate time series, which is the argument of this article. So if we refers to a trends in flood flows: Y would be streamflow, X would be the precipitation amount, and R would be called the precipitation-adjusted flow (the duration of precipitation used must be appropriate to the flow variable under consideration. For example, if Y is the annual flood peak from a 25 km² basin then X might be the 1-hour maximum rainfall, whereas if Y is the annual flood peak for a 25000 km² basin then X might be the 24-hour maximum rainfall).

Mann-Kendall test is related with a statistical hypothesis testing. The null hypotesis $H_0$ is that there is no trend in the time series. The three alternative hypotesis are that there is a negative trend, a non-null trend, or a positive trend.

|  | True Situation | |
|---|---|---|
| Decision | No trend. $H_0$ true. | Trend exists. $H_0$ false. |
| Fail to reject $H_0$. "No trend" | Probability = $1-\alpha$ | (Type II error) $\beta$ |
| Reject $H_0$. "Trend" | (Type I error) significance level $\alpha$ | (Power) $1-\beta$ |

Probabilities associated with possible outcomes of a trend test.
$\alpha$ = Prob (reject $H_0 | H_0$ true)   and   $1-\beta$ = Prob (reject $H_0 | H_0$ false)

The hypothesis testing is performed according to a predetermined level of significance, in a manner that the test assumes meaning or not. A p-value less than the predetermined significance indicates that the null hypothesis is rejected, i.e. that there is a trend in our time series.

## 5. AUTOREGRESSION.

### 5.1 BASIS.

We state that data at particular point of time are (probably highly) correlated with the value that precede and/or succeed.
The creation of an autoregressive model generates a new predictor variable by using the Y variable lagged one or more periods.

$$y_t = f(y_{t-1}, y_{t-2}, ..., y_{t-p}, \varepsilon_t) \quad (5.1)$$

Dependent variable is a function of itself at the previous moment of period or time.
The most often seen form of the equation is a linear form:

$$y_t = b_0 + \sum_{i=1}^{p} b_i y_{t-i} + e_t \quad (5.2)$$

where:
$y_t$ the dependent variable values at the moment *t*,
$y_{t-i}$ (i = 1, 2, ..., p) the dependent variable values at the moment *t-i*,
$b_0$, $b_i$ (i=1,..., p) regression coefficient,
p autoregression rank,
$e_t$ disturbance term.

$$b = \begin{bmatrix} b_0 \\ b_1 \\ \vdots \\ b_p \end{bmatrix} \quad y = \begin{bmatrix} y_{p+1} \\ y_{p+2} \\ \vdots \\ y_n \end{bmatrix} \quad X = \begin{bmatrix} 1 & y_p & y_{p-1} & \cdots & y_1 \\ 1 & y_{p+1} & y_p & & y_2 \\ \vdots & \vdots & \vdots & & \vdots \\ 1 & y_{n-1} & y_{n-2} & & y_{n-p} \end{bmatrix} \quad (5.3)$$

A first-order autoregressive model is concerned with only the correlation between consecutive values in a series.

$$y_t = b_0 + b_1 y_{t-1} + e_t \quad (5.4)$$

A second-order autoregressive model considers the effect of relationship between consecutive values in a series as well as the correlation between values two periods apart.

$$y_t = b_0 + b_1 y_{t-1} + b_2 y_{t-2} + e_t \quad (5.5)$$

The selection of an appropriate autoregressive model is not an easy task.

Once a model is selected and OLS method is used to obtain estimates of the parameters, the next step would be to eliminate those parameters which do not contribute significantly.

$$H_0; \beta_p = 0 \qquad (5.6)$$

(The highest-order parameter does not contribute to the prediction of Yt)

$$H_1; \beta_p \neq 0 \qquad (5.7)$$

(The highest-order parameter is significantly meaningful)

$$Z = \frac{b_p}{S(b_p)} \qquad (5.8)$$

using an alpha level of significance, the decision rule is to reject $H_0$ if

$$Z > Z_\alpha \qquad (5.9)$$

or if

$$-Z < -Z_\alpha \qquad (5.10)$$

and not to reject $H_0$ if

$$-Z_\alpha \leq Z \leq Z_\alpha \qquad (5.11)$$

Here a short table with values of $Z_\alpha$

| |
|---|
| $Z_{0,1} = 1,645$ |
| $Z_{0,05} = 1,960$ |
| $Z_{0,02} = 2,236$ |
| $Z_{0,01} = 2,576$ |
| $Z_{0,001} = 3,291$ |

If the null hypothesis is NOT rejected we may conclude that the selected model contains too many estimated parameters. The highest-order term then be deleted an a new autoregressive model would be obtained through least-squares regression. A test of the hypothesis that the "new" highest-order term is 0 would then be repeated.
This testing and modeling procedure continues until we reject $H_0$. When this occurs, we know that our highest-order parameter is significant and we are ready to use this model.

5.2     CASE STUDY.

5.2.1   THE MODEL.

Refering to the paper wrote about extreme events in Italy (Climate events and insurance demand - the effect of potentially catastrophic events on insurance demand in Italy, G. Rosso, A. Chieppa, A. Ricca, with C. D. Pronzato and I. Pecetto, 2014).

We used a complete database with daily climate observations. The database was processed and modified in order to have: first, a smaller monthly database, second, a set with the only extreme events.

This database is shown in tab. 5.2.1.1. It represent extrem precipitation events in last eleven years in Italy. The first event assume the "month" number 1, so that di other events have a month numbered with a properly "jump", usefull to draw correctly in a chart, and to calculate a frequency and the "distance" between each single event.

The tab. 5.2.1.1 show also the amount of precipitation in millimeters. You can note that there are no events with precipitations less than 100 mm. This was the threshold assumed in the past research. Substantially, in that research the threashold was linear, as shown in fig. 3.4. Because the study of the regression, seems to be incompatible with charts like 3.5 or 3.6.

This is the minimum dataset required for an autoregression analysis.

| month | precip mm | month | precip mm |
|---|---|---|---|
| 1 | 200 | 94 | 800 |
| 8 | 396 | 95 | 517 |
| 9 | 280 | 99 | 363 |
| 24 | 517 | 102 | 195,2 |
| 40 | 150 | 106 | 542 |
| 43 | 203 | 107 | 927 |
| 45 | 154 | 119 | 899 |
| 58 | 200 | 128 | 405 |
| 65 | 312 | 130 | 601 |
| 70 | 372 | 131 | 400 |
| 79 | 195,2 | 133 | 570 |
| 82 | 225 | 137 | 130 |
| 84 | 381 | 139 | 300 |
| 86 | 355 | 140 | 520 |
| 93 | 230 | 141 | 500 |
| >>>>> | | 142 | 1355 |

*Tab. 5.2.1.1*

As said, you can note that the column "month" jumps. In Our case this is not a problem because the autoregression model correltes the observations among them, properly shiftet for the number of periods we want to analize. But there could be cases in which also the"empty" observation plays an important role. This could be for reinsurance, where the threshold represent the excess amount and the zero items could have significance if the Company payed nothing). We could explain this approach with a very simple example.

Let the series x(t) be observed, with t = (1, … , 12)

| | |
|---|---|
| 1 | 200 |
| 2 | 30 |
| 3 | 40 |
| 4 | 120 |
| 5 | 80 |
| 6 | 110 |
| 7 | 180 |
| 8 | 55 |
| 9 | 190 |
| 10 | 110 |
| 11 | 20 |
| 12 | 110 |

*Tab. 5.2.1.2*

Using the POT peaks over threshold method on order to save the only observations exceeding 100 of value, the result is shown in the following Tab. 5.2.1.3.

| | |
|---|---|
| 1 | 200 |
| 2 | – |
| 3 | – |
| 4 | 120 |
| 5 | – |
| 6 | 110 |
| 7 | 180 |
| 8 | – |
| 9 | 190 |
| 10 | 110 |
| 11 | – |
| 12 | 110 |

*Tab. 5.2.1.3*

As said, this dataset can be considerend in two different ways:

**tab A**

| | |
|---|---|
| 1 | 200 |
| 4 | 120 |
| 6 | 110 |
| 7 | 180 |
| 9 | 190 |
| 10 | 110 |
| 12 | 110 |

**tab B**

| | |
|---|---|
| 1 | 200 |
| 2 | 0 |
| 3 | 0 |
| 4 | 120 |
| 5 | 0 |
| 6 | 110 |
| 7 | 180 |
| 8 | 0 |
| 9 | 190 |
| 10 | 110 |
| 11 | 0 |
| 12 | 110 |

*Tab. 5.2.1.4 – 5.2.1.4.bis*

These concepts are well described by Stuart Coles, from Bristol University.
Fig. 5.2.1.1. shows an example of chart that represent precipitations.

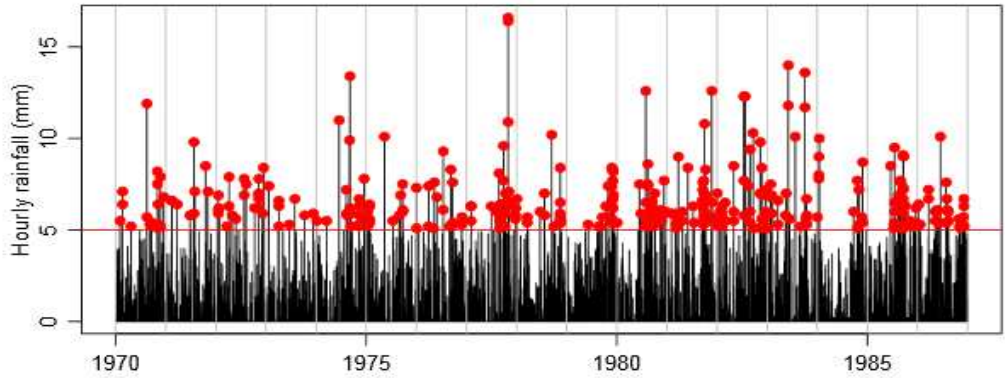

*Fig. 5.2.1.1 Source: S. Coles, Bristol University*

Events under the threshold are dropped in Fig. 5.2.1.2.

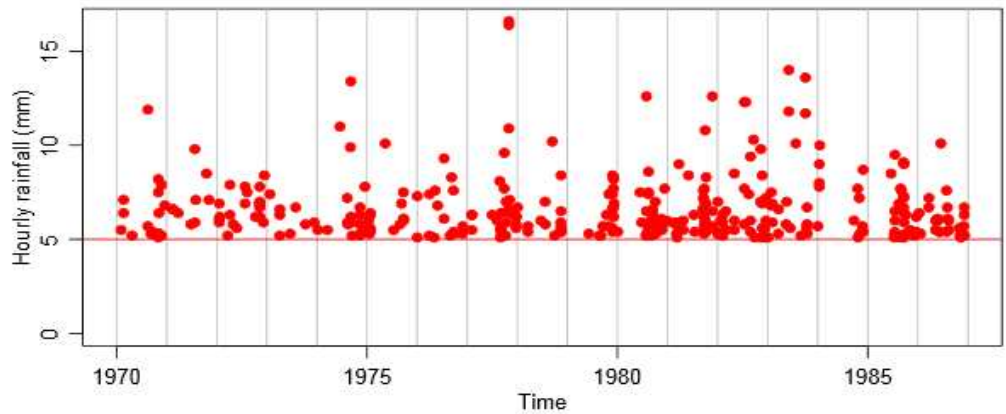

*Fig. 5.2.1.2: S. Coles, Bristol University*

What we need is to consider these events consequential among them.
The Tab A is good for our scopes, and easy treatable with the Autoregression method. With this dataset we can draw a scatterplot and we can conduct a prior analysis in order to evidence if a trend is present in our time series.
The scatterplot of data shows an increasing trend (Fig. 5.2.1.3).
The regression was calculated as said in par. 4, and the equation of the regression line is

$$y = 14,78x + 189,14 \qquad (5.2.1.1)$$

One of the condition in order to make a good autoregression analysis is that we need to detrend our series; one of the reason is to be able to obtain meaningful sample statistics such as means, variances, and correlations with other variables. Such statistics are useful as descriptors of future behavior only if the series is stationary. For example, if the series is consistently increasing over time, the sample mean and variance will grow with the size of the sample, and they will always underestimate the mean and variance in future periods. And if the mean and variance of a series are not well-defined, then neither are its correlations with other variables. For this reason you should be cautious about trying to extrapolate regression models fitted to nonstationary data. But many analysis uses the original time series not detrended. The reason could be found in the values contained in the database, which propriety and behaviour drives to prefer this kind of analysis. We will use the autocorrelation for both, trended and detrended time series; and we will compare the results.

Many alternative methods are available for detrending. Simple linear trend in mean can be removed by subtracting the least-squares-fit straight line calculated with regression.

Therefore the equation (5.2.1.1) is usefull to do detrending procedure. Many statistical software do this operation automatically, but even in Excel a simply formula can resolve this problem. For time series trends, Excel treats the equation as a function of observation number, so that the first observation is 1, the second is 2; and so on. It is necessary to add a column with these observation numbers to support the trend calculations.

Figure 5.2.1.3 shows the method of calculation, with the formula used for detrending. It's easy to recognize into the formula the parameters of the regression line. The new detrended series is then shifted for three periods. These data will be used in autoregression analysis.

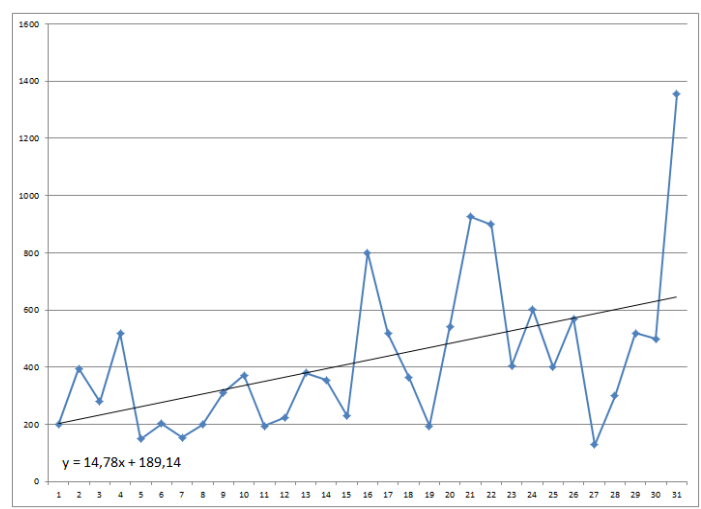

*Fig. 5.2.1.3*

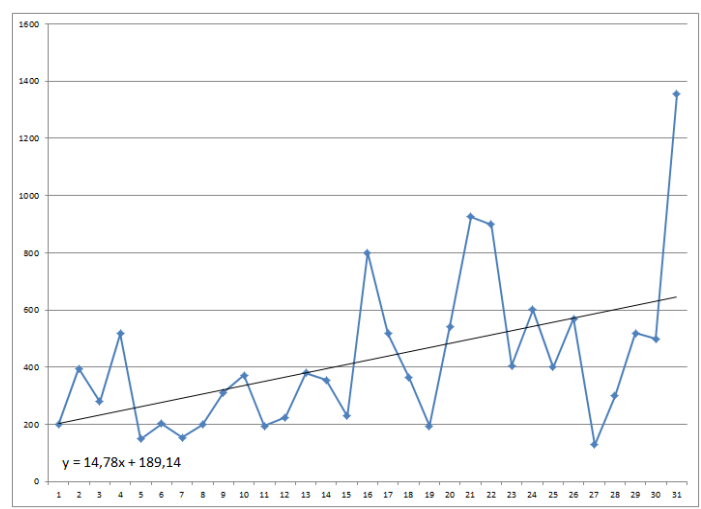

*Fig. 5.2.1.4*

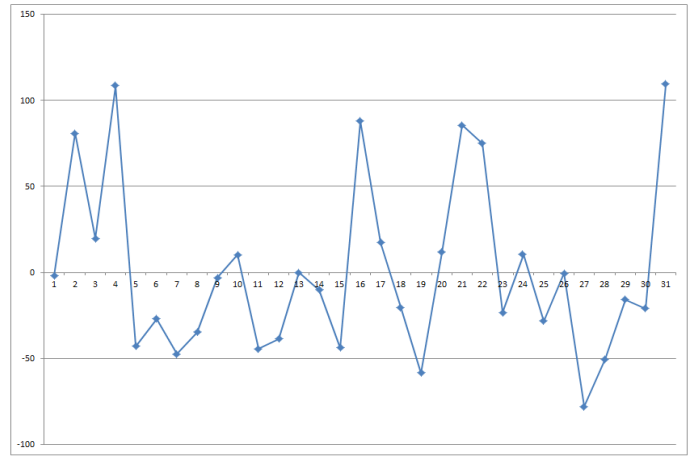

*Fig. 5.2.1.5*

The scatterplot of the detrended series is pictured in figure 5.2.1.5.
The first step to use the autoregression model is to replicate all the independent variables, one for each period you think usefull for the analysis. The key is to copy data near the original data, but one-period shifted. As said, this operation can be replicated even if you have di amount of shifted periods you want to analyze.
In Fig. 5.2.1.6 you can see the result: the Tab 5.2.1.1 integrated with the series detrended from the Fig. 5.2.1.5, plus the shifting action to generate a numbers of other variables usefull to perform an adeguate autoregressive analysis. The data at the bottom of the table will be never used. The data at the top will be used in correletion to the period of the autoregression you need to perform.
In this model are used detrended data calculated (Fig. 5.2.1.4), but the procedure is the same even with original data not detrended.

| | A | B | C | D | E | F | G |
|---|---|---|---|---|---|---|---|
| 1 | month | precip mm | # | detr | X1 | X2 | X3 |
| 2 | 1 | 200 | 1 | -1,922 | | | |
| 3 | 8 | 396 | 2 | 81,07 | -1,922 | | |
| 4 | 9 | 280 | 3 | 19,925 | 81,07 | -1,922 | |
| 5 | 24 | 517 | 4 | 108,25 | 19,925 | 81,07 | -1,922 |
| 6 | 40 | 150 | 5 | -42,97 | 108,25 | 19,925 | 81,07 |
| 7 | 43 | 203 | 6 | -26,93 | -42,97 | 108,25 | 19,925 |
| 8 | 45 | 154 | 7 | -47,37 | -26,93 | -42,97 | 108,25 |
| 9 | 58 | 200 | 8 | -34,93 | -47,37 | -26,93 | -42,97 |
| 10 | 65 | 312 | 9 | -3,154 | -34,93 | -47,37 | -26,93 |
| 11 | 70 | 372 | 10 | 10,405 | -3,154 | -34,93 | -47,37 |
| 12 | 79 | 195,2 | 11 | -44,5 | 10,405 | -3,154 | -34,93 |
| 13 | 82 | 225 | 12 | -38,61 | -44,5 | 10,405 | -3,154 |
| 14 | 84 | 381 | 13 | -0,073 | -38,61 | -44,5 | 10,405 |
| 15 | 86 | 355 | 14 | -10,37 | -0,073 | -38,61 | -44,5 |
| 16 | 93 | 230 | 15 | -44,02 | -10,37 | -0,073 | -38,61 |
| 17 | 94 | 800 | 16 | 87,961 | -44,02 | -10,37 | -0,073 |
| 18 | 95 | 517 | 17 | 17,393 | 87,961 | -44,02 | -10,37 |
| 19 | 99 | 363 | 18 | -20,25 | 17,393 | 87,961 | -44,02 |
| 20 | 102 | 195,2 | 19 | -58,46 | -20,25 | 17,393 | 87,961 |
| 21 | 106 | 542 | 20 | 11,813 | -58,46 | -20,25 | 17,393 |
| 22 | 107 | 927 | 21 | 85,578 | 11,813 | -58,46 | -20,25 |
| 23 | 119 | 899 | 22 | 74,801 | 85,578 | 11,813 | -58,46 |
| 24 | 128 | 405 | 23 | -23,45 | 74,801 | 85,578 | 11,813 |
| 25 | 130 | 601 | 24 | 10,506 | -23,45 | 74,801 | 85,578 |
| 26 | 131 | 400 | 25 | -28,4 | 10,506 | -23,45 | 74,801 |
| 27 | 133 | 570 | 26 | -0,596 | -28,4 | 10,506 | -23,45 |
| 28 | 137 | 130 | 27 | -77,9 | -0,596 | -28,4 | 10,506 |
| 29 | 139 | 300 | 28 | -50,25 | -77,9 | -0,596 | -28,4 |
| 30 | 140 | 520 | 29 | -15,82 | -50,25 | -77,9 | -0,596 |
| 31 | 141 | 500 | 30 | -20,95 | -15,82 | -50,25 | -77,9 |
| 32 | 142 | 1355 | 31 | 109,32 | -20,95 | -15,82 | -50,25 |
| 33 | | | | | 109,32 | -20,95 | -15,82 |
| 34 | | | | | | 109,32 | -20,95 |
| 35 | | | | | | | 109,32 |
| 36 | | | | | | | |

*Fig. 5.2.1.6*

Analysis for one period with the variable "detr" as dependent and X1 as independent.

Note that because the shifting, the first value of "detr" is not considered.

The regression output is typical. All outputs are showed as attachments, both detrended and trended series.

The way to add and examine more periods is easy and is showed in Fig. 5.2.1.8.

At the top of the Excel sheet are data that will be used only if needed in relationship with the regressors at the right of the original data. Every single period *i* of the autoregressive method AR(*i*) it's none other than a single multiregression, with the exception on the first one that is a single regression with a dependent and an independent variable.

*Fig. 5.2.1.7*          *Fig. 5.2.1.8*

About the method to read regression output, here are few indications.

$R^2$ is between 0 and 1 and indicates the amount of variation of $y_i$ around $y_{bar}$ (its mean) that is explained by the regressors.

The ANOVA (analysis of variance) table splits the sum of squares into its components:

*Total sums of squares = Residual (or error) sum of squares + Regression (or explained) sum of squares.*

Thus

$$\Sigma_i (y_i - ybar)^2 = \Sigma_i (y_i - yhat_i)^2 + \Sigma_i (yhat_i - ybar)^2 \qquad (5.2.1.2)$$

where yhat$_i$ is the value of y$_i$ predicted from the regression line and ybar is the sample mean of y.

$$R^2 = 1 - Residual\ SS\ /\ Total\ SS \quad \text{(general formula for R}^2\text{)} \quad (5.2.1.3)$$

F gives the overall F-test of H0: β$_j$ = 0 versus Ha: at least one of β$_j$ does not equal zero. Excel computes F this as

$$F = [Regression\ SS/(k-1)]\ /\ [Residual\ SS/(n-k)] \quad (5.2.1.4)$$

Significance F has the associated P-value. If value > 0.05, we do not reject H0 at signficance level 0.05. Significance F in general has a related Excel formula that is FINV(F, k-1, n-k) where k is the number of regressors including hte intercept.

- Column "Coefficient" gives the least squares estimates of β$_j$.
- Column "Standard error" gives the standard errors (i.e.the estimated standard deviation) of the least squares estimates b$_j$ of β$_j$.
- Column "t Stat" gives the computed t-statistic for H0: β$_j$ = 0 against Ha: β$_j$ ≠ 0.
- This is the coefficient divided by the standard error. It is compared to a t with (n-k) degrees of. N is the number of observations and k tha number of regressors.
- Column "P-value" gives the p-value for test of H0: β$_j$ = 0 against Ha: β$_j$ ≠ 0.
- This equals the Pr{|t| > t-Stat}where t is a t-distributed random variable with n-k degrees of freedom and t-Stat is the computed value of the t-statistic given in the previous column. Note that this p-value is for a two-sided test. For a one-sided test divide this p-value by 2 (also checking the sign of the t-Stat).
- Columns "Lower 95%" and "Upper 95%" values define a 95% confidence interval for β$_j$.

We used Excel to perform the analysis, because Excel has a very large diffusion, a lot of tools designed for it are available, frequently for free, and costs for the program are very light if rapported with others statistical programs, expecially for students. But excel own particulary specs that is important to know:
- Excel restricts the number of regressors (only up to 16 regressors);
- Excel requires that all the regressor variables be in adjoining columns. You may need to move columns to ensure this;
- Excel standard errors and t-statistics and p-values are based on the assumption that the error is independent with constant variance (homoskedastic);
- Excel does not provide alternatives, such asheteroskedastic-robust or autocorrelation-robust standard errors and t-statistics and p-values;
- more specialized software are STATA, EVIEWS, SAS, LIMDEP, PC-TSP, and many others.

Now, let's see the output of detrended AR(1).

OUTPUT RIEPILOGO

| Statistica della regressione | |
|---|---|
| R multiplo | 0,148665849 |
| R al quadrato | 0,022101535 |
| R al quadrato corretto | -0,012823411 |
| Errore standard | 52,17970381 |
| Osservazioni | 30 |

ANALISI VARIANZA

| | gdl | SQ | MQ | F | Significatività F |
|---|---|---|---|---|---|
| Regressione | 1 | 1723,018403 | 1723,018403 | 0,632829472 | 0,433012 |
| Residuo | 28 | 76236,2017 | 2722,721489 | | |
| Totale | 29 | 77959,22011 | | | |

| | Coefficienti | Errore standard | Stat t | Valore di significatività | Inferiore 95% | Superiore 95% | Inferiore 95,0% | Superiore 95,0% |
|---|---|---|---|---|---|---|---|---|
| Intercetta | 1,382648743 | 9,543368813 | 0,144880573 | 0,885843027 | -18,1661 | 20,931354 | -18,16606 | 20,931354 |
| Variabile X 1 | 0,161812235 | 0,203407989 | 0,795505797 | 0,433011776 | -0,25485 | 0,5784746 | -0,25485 | 0,5784746 |

*Fig. 5.2.1.9*

Standard deviation of data is 50,9 (form Descriptive Statistics). The standard error of the autoregression (52,17) and the t Stat (0,79) seems to tell us that the data are not so lagged as we expected. Also autoregressions with more periods drives to the same conclusions.

OUTPUT RIEPILOGO

| Statistica della regressione | |
|---|---|
| R multiplo | 0,15496 |
| R al quadrato | 0,024012602 |
| R al quadrato corretto | -0,051063352 |
| Errore standard | 51,74028171 |
| Osservazioni | 29 |

ANALISI VARIANZA

| | gdl | SQ | MQ | F | Significatività F |
|---|---|---|---|---|---|
| Regressione | 2 | 1712,481664 | 856,2408322 | 0,319844109 | 0,729079 |
| Residuo | 26 | 69603,47554 | 2677,056752 | | |
| Totale | 28 | 71315,95721 | | | |

| | Coefficienti | Errore standard | Stat t | Valore di significatività | Inferiore 95% | Superiore 95% | Inferiore 95,0% | Superiore 95,0% |
|---|---|---|---|---|---|---|---|---|
| Intercetta | -1,401310389 | 9,630257288 | -0,14551121 | 0,885429498 | -21,1966 | 18,393967 | -21,19659 | 18,393967 |
| Variabile X 1 | 0,163426393 | 0,205573011 | 0,794979805 | 0,433822783 | -0,25913 | 0,5859878 | -0,259135 | 0,5859878 |
| Variabile X 2 | -0,013931734 | 0,206107918 | -0,067594365 | 0,946625678 | -0,43759 | 0,4097292 | -0,437593 | 0,4097292 |

*Fig. 5.2.1.10*

OUTPUT RIEPILOGO

| Statistica della regressione | |
|---|---|
| R multiplo | 0,327250905 |
| R al quadrato | 0,107093155 |
| R al quadrato corretto | -0,004520201 |
| Errore standard | 51,3325957 |
| Osservazioni | 28 |

ANALISI VARIANZA

| | gdl | SQ | MQ | F | Significatività F |
|---|---|---|---|---|---|
| Regressione | 3 | 7584,959273 | 2528,319758 | 0,959501256 | 0,427849 |
| Residuo | 24 | 63240,84915 | 2635,035381 | | |
| Totale | 27 | 70825,80842 | | | |

| | Coefficienti | Errore standard | Stat t | Valore di significatività | Inferiore 95% | Superiore 95% | Inferiore 95,0% | Superiore 95,0% |
|---|---|---|---|---|---|---|---|---|
| Intercetta | -2,141059352 | 9,783612282 | -0,218841394 | 0,828623951 | -22,3334 | 18,051324 | -22,33344 | 18,051324 |
| Variabile X 1 | 0,152825569 | 0,216790797 | 0,704944908 | 0,487630083 | -0,29461 | 0,6002598 | -0,294609 | 0,6002598 |
| Variabile X 2 | 0,048112684 | 0,208573612 | 0,230674838 | 0,81952248 | -0,38236 | 0,4785875 | -0,382362 | 0,4785875 |
| Variabile X 3 | -0,316125243 | 0,20466228 | -1,544618984 | 0,135523417 | -0,73853 | 0,1062769 | -0,738527 | 0,1062769 |

*Fig. 5.2.1.11*

The assumption was that climate extreme events can have a relationship, a lagged effect, because there is an element, or more than one, that influence results, so that the result of one time period tends to spill over into the next period or periods. We know that probably exogenous effects are influencing frequency and intensity of events, and that seems to be real an accumulation situation. But the analysis with a detrended series is so borderline, even with a conclusion to a ngative direction, that probably the "noise" that define the trend has characteristics so particulary that a detrending action is no needed.

In effect the autoregression analysis performed with the original data give other results.

Standard deviation of the descriptive statisticis (271,6) is grater than standard error we get with autoregression with one period. So the original data seems to be lagged.

OUTPUT RIEPILOGO

| Statistica della regressione | |
|---|---|
| R multiplo | 0,32803921 |
| R al quadrato | 0,10760973 |
| R al quadrato corretto | 0,07573865 |
| Errore standard | 262,491897 |
| Osservazioni | 30 |

ANALISI VARIANZA

| | gdl | SQ | MQ | F | Significatività F |
|---|---|---|---|---|---|
| Regressione | 1 | 232641,1 | 232641,148 | 3,376406518 | 0,07677 |
| Residuo | 28 | 1929256 | 68901,9959 | | |
| Totale | 29 | 2161897 | | | |

| | Coefficienti | Errore standard | Stat t | Valore di significatività | Inferiore 95% | Superiore 95% | Inferiore 95,0% | Superiore 95,0% |
|---|---|---|---|---|---|---|---|---|
| Intercetta | 267,592408 | 102,0505 | 2,62215706 | 0,013972464 | 58,55146 | 476,63336 | 58,551459 | 476,63336 |
| Variabile X 1 | 0,41949996 | 0,228299 | 1,83750007 | 0,0767705 | -0,04815 | 0,8871498 | -0,04815 | 0,8871498 |

*Fig. 5.2. 1.12*

We should have a proof with the correlation analysis.

**detrended**

|   | data | X1 |
|---|---|---|
| data | 1 | |
| X1 | 0,148665849 | 1 |

**trended**

|   | data | X1 |
|---|---|---|
| data | 1 | |
| X1 | 0,32803921 | 1 |

*Fig. 5.2.1.13*

In both the correlation is positive, as expected. As in the detrended data value is less than 0.3, the correlation is weak. But in the correlation calculated with original data the correlation is just a little bit better, a moderate correlation that follow the results of autoregression. With two and three periods, also the autoregression with original data tell us that the relationship between the events semms to be non existent.

**OUTPUT RIEPILOGO**

| Statistica della regressione | |
|---|---|
| R multiplo | 0,33304455 |
| R al quadrato | 0,11091867 |
| R al quadrato corretto | 0,0425278 |
| Errore standard | 271,805443 |
| Osservazioni | 29 |

**ANALISI VARIANZA**

|   | gdl | SQ | MQ | F | Significatività F |
|---|---|---|---|---|---|
| Regressione | 2 | 239636,4 | 119818,205 | 1,621834416 | 0,21689 |
| Residuo | 26 | 1920833 | 73878,1987 | | |
| Totale | 28 | 2160470 | | | |

|   | Coefficienti | Errore standard | Stat t | Valore di significatività | Inferiore 95% | Superiore 95% | Inferiore 95,0% | Superiore 95,0% |
|---|---|---|---|---|---|---|---|---|
| Intercetta | 244,810699 | 125,7253 | 1,94718708 | 0,062389478 | -13,6214 | 503,24278 | -13,62138 | 503,24278 |
| Variabile X 1 | 0,40025971 | 0,256212 | 1,5622227 | 0,130326579 | -0,12639 | 0,9269104 | -0,126391 | 0,9269104 |
| Variabile X 2 | 0,07408742 | 0,253488 | 0,29227157 | 0,772398608 | -0,44697 | 0,59514 | -0,446965 | 0,59514 |

*Fig. 5.2.1.14*

**OUTPUT RIEPILOGO**

| Statistica della regressione | |
|---|---|
| R multiplo | 0,3494325 |
| R al quadrato | 0,12210307 |
| R al quadrato corretto | 0,01236595 |
| Errore standard | 279,507447 |
| Osservazioni | 28 |

**ANALISI VARIANZA**

|   | gdl | SQ | MQ | F | Significatività F |
|---|---|---|---|---|---|
| Regressione | 3 | 260784,1 | 86928,0242 | 1,112687067 | 0,363433 |
| Residuo | 24 | 1874986 | 78124,4132 | | |
| Totale | 27 | 2135770 | | | |

|   | Coefficienti | Errore standard | Stat t | Valore di significatività | Inferiore 95% | Superiore 95% | Inferiore 95,0% | Superiore 95,0% |
|---|---|---|---|---|---|---|---|---|
| Intercetta | 293,363292 | 146,471 | 2,00287647 | 0,056609992 | -8,93797 | 595,66455 | -8,937966 | 595,66455 |
| Variabile X 1 | 0,40810533 | 0,263935 | 1,54623378 | 0,135133901 | -0,13663 | 0,9528405 | -0,13663 | 0,9528405 |
| Variabile X 2 | 0,10155194 | 0,280089 | 0,36257076 | 0,720098983 | -0,47652 | 0,6796264 | -0,476523 | 0,6796264 |
| Variabile X 3 | -0,1493143 | 0,261725 | -0,5705011 | 0,573640953 | -0,68949 | 0,3908592 | -0,689488 | 0,3908592 |

*Fig. 5.2.1.15*

### 5.2.2 TESTING THE MODEL.

We have demonstrated that a the original dataset has a lightly autocorrelation with one period. Nothing concern other periods and the detrended series.
The model building need another important step: the validation.
Model validation is a very important step in the model building sequence. Often the validation of a model seems to consist of nothing more than quoting the $R^2$ statistic from the fit. But we have seen that in our specific case $R^2$ is particularly low. This is not necessarily bad because $R^2$ value does not guarantee that the model fits the data well. So, at this point we need to do a graphical residual analysis.
The difference between the observed value of the dependent variable ($y$) and the predicted value ($\hat{y}$) is called the residual ($e$). Each data point has one residual.
Residual = Observed value - Predicted value

$$e = y - \hat{y} \qquad (5.2.2.1)$$

The residual plot shows the residuals on the vertical axis and the independent variable on the horizontal axis. If the points in a residual plot are randomly dispersed around the horizontal axis, a linear regression model is appropriate for the data; otherwise, a non-linear model is more appropriate.
The residual plot of our model is shown in the figure below (Fig. 5.2.2.1).

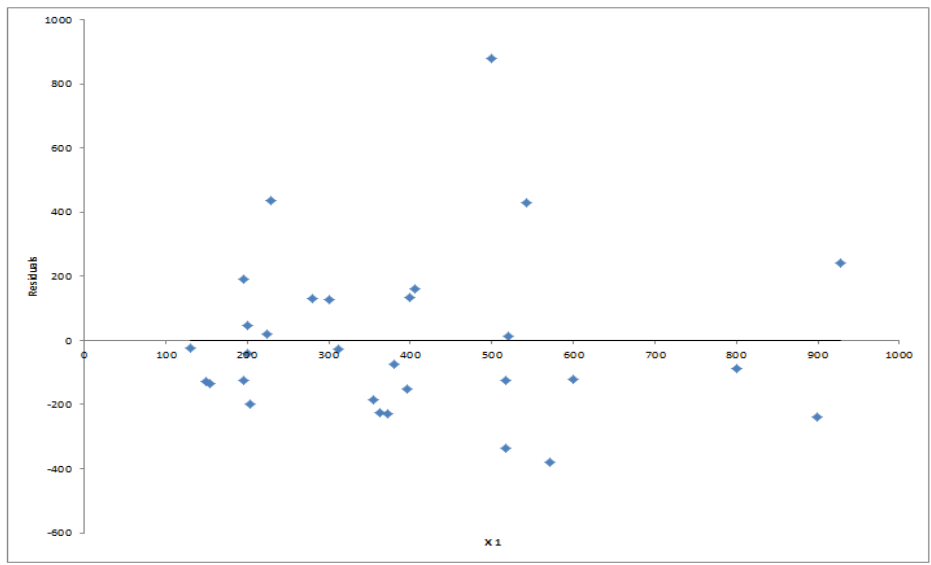

*Fig. 5.2.2.1*

The residual plot shows a fairly random. This random pattern indicates that a linear model provides a decent fit to the data. The residuals should not be either systematically high or low. So, the residuals should be centered on zero throughout the range of fitted values. In other words, the model is correct on average for all fitted values.
Nontheless we have a situation particulary good, and the residual plot seems to confirm that the model describe farly well our time series, the plot show in the middle of X axis a point particulary high, that seems to be over the average and is distinctly over the other values.

Displaying the data related to this point we could see that the value is 500 and the difference (residual) is 877.65 (Fig. 5.2.2.2 and Fig. 5.2.2.3).

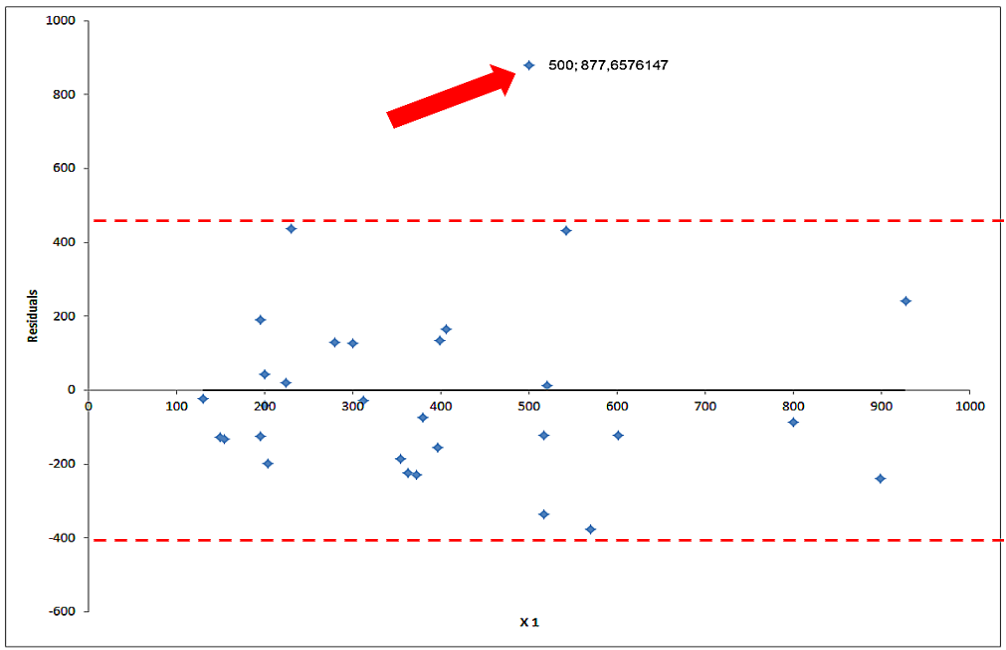

*Fig. 5.2.2.3*

| Osservazione | Y prevista | Residui | Residui standard | Percentile | Y |
|---|---|---|---|---|---|
| 1 | 351,4924 | 44,50760128 | 0,17255926 | 1,666666667 | 130 |
| 2 | 433,71439 | -153,7143899 | -0,595962053 | 5 | 150 |
| 3 | 385,0524 | 131,9476049 | 0,511570618 | 8,333333333 | 154 |
| 4 | 484,47388 | -334,4738845 | -1,296779976 | 11,66666667 | 195,2 |
| 5 | 330,5174 | -127,517401 | -0,49439439 | 15 | 195,2 |
| 6 | 352,7509 | -198,7508986 | -0,770571927 | 18,33333333 | 200 |
| 7 | 332,1954 | -132,1954008 | -0,512531341 | 21,66666667 | 203 |
| 8 | 351,4924 | -39,49239872 | -0,153114949 | 25 | 225 |
| 9 | 398,47639 | -26,4763937 | -0,102650936 | 28,33333333 | 230 |
| 10 | 423,64639 | -228,446391 | -0,885703546 | 31,66666667 | 280 |
| 11 | 349,4788 | -124,4787989 | -0,482613506 | 35 | 300 |
| 12 | 361,9799 | 19,0201024 | 0,073742343 | 38,33333333 | 312 |
| 13 | 427,42189 | -72,42189061 | -0,280785024 | 41,66666667 | 355 |
| 14 | 416,51489 | -186,5148918 | -0,723132024 | 45 | 363 |
| 15 | 364,0774 | 435,9226026 | 1,690104156 | 48,33333333 | 372 |
| 16 | 603,19237 | -86,19237184 | -0,334174197 | 51,66666667 | 381 |
| 17 | 484,47388 | -121,4738845 | -0,47096323 | 55 | 396 |
| 18 | 419,87089 | -224,6708914 | -0,871065655 | 58,33333333 | 400 |
| 19 | 349,4788 | 192,5212011 | 0,746418929 | 61,66666667 | 405 |
| 20 | 494,96138 | 432,0386166 | 1,675045655 | 65 | 500 |
| 21 | 656,46887 | 242,5311339 | 0,940311135 | 68,33333333 | 517 |
| 22 | 644,72287 | -239,7228674 | -0,929423279 | 71,66666667 | 517 |
| 23 | 437,48989 | 163,5101105 | 0,633940786 | 75 | 520 |
| 24 | 519,71188 | -119,7118808 | -0,464131811 | 78,33333333 | 542 |
| 25 | 435,39239 | 134,6076102 | 0,521883656 | 81,66666667 | 570 |
| 26 | 506,70738 | -376,7073821 | -1,460522369 | 85 | 601 |
| 27 | 322,1274 | -22,12740186 | -0,085789573 | 88,33333333 | 800 |
| 28 | 393,44239 | 126,5576058 | 0,490673193 | 91,66666667 | 899 |
| 29 | 485,73238 | 14,26761562 | 0,0553166 | 95 | 927 |
| 30 | 477,34239 | 877,6576147 | 3,402743454 | 98,33333333 | 1355 |

*Fig. 5.2.2.2*

This point is refered to the last observation with 1355 in the 142nd month. After the procedure of identification of peaks, the residual plot help us to find outliers. This event is clearly an outliers. As wrote in a previous paper (G. Rosso, Outilers emphasis on cluster analysis - the use of squared Euclidean distance and fuzzy clustering to detect outliers in a dataset, 2014) outliers can became a precious source of information. In our case we can note that the anomalous event id the last of our time series, and this bring us to the need to persist in carefully observation of events because probably in a close future we could have other anomalous events.

## 6 CONCLUSIONS.

Autocorrelation, trend, autoregression. These items are strictly lagged among them. When we try to understand events, frequently time series that describe them have caracteristics of autocorrelation. This means that every single events could be influenced by the previous event, and often not with the only one previous, but with many.

As a natural continuation of the research "Climate events and insurance demand - The effect of potentially catastrophic events on insurance demand in Italy" we tried to answer to these questions: these kind of time series have caracteristic of autocorrelation? How can we manage this events? There is the opportunity to assess them correctly and to do predictions?

The answer to the first question is: yes, frequently risk time series, and climate risk time series in particulary, have events correlated among them.

The answer to the second question was explained in this article. We suggested to manage first the amount of the events, because probably not the entire time series could be usefull for analysis. So if only events exceeding a threshold are usefull for the analysis, there are methods dedicated to this work of selection. After that, we suggest to consider that events could be correlated, and that a kind of "noise" can influence the trend of the time series. We can image this "noise" as a persistence of any kind.

What supposed often is difficult to demonstrate. The time series object of study must be treated carefully and every result must be interpreted. In our example, autoregression was calculated with the data of the citated research. The scatterplot shows evident fluctuations especially in the last period, however the trend is clear. We know that climate time series are affected by a persistence noise. The world climate is changing, and related risks too. The change seems to be very fast, progressive and exponential increasing. The effect of our anthropogenic activities are subjected to accumulation. This is the "noise" we are speaking. And the increasing trend seems to be in that direction. But if we see the results of the autoregression calculated with our climate data, this assumption is not evident. The time series was detrended before the autoregression calculus, but results tell us that the original time series behave than the detrended series, and only for one period. Results of the autoregression calculated (both on the detrended time series and the original one) seem to suggest that the only effect that evoke the presence of this noise is really the trend. Besides the strong fluctuation of the data values produces outputs with feeble significance of the R squared. But stronger evidence are in the residual analysis. This kind of analysis help to verify the goodness of the model, but also help to identify outliers present in time series.

Definitively we can assert that in a climate time series is present a noise that drive to a correlation between periods. The process proposed help to define if the series must be detrended, how many periods are lagged, if model is good enough, if there are outliers in the series and what is their magnitude. The process is long, but supply a lot of informations usefull to provide directions about the analysis.

ACKNOWLEDGEMENTS.

*Matteo Barigozzi*, LSE London School of Economics and Political Science;
*Esterina Masiello*, UCBL Claude Bernard University Lyon 1 - ISFA Institut de science financière et d'assurances;
*Keith Porter,* University of Colorado Boulder and SPA Risk LLC, Denver CO USA.